\documentclass[a4paper,oneside,12pt]{article}
\pdfoutput=1 

\usepackage[textwidth=17cm,textheight=24cm]{geometry}
\usepackage{hyperref}
\usepackage{authblk}
\usepackage[pdftex]{graphicx}
\usepackage[font=small]{caption}   

\usepackage{placeins}

\providecommand{\keywords}[1]{\textbf{\textit{Keywords---}} #1}

\begin{document}

This is an author-created, un-copyedited version of an article accepted for publication in JINST. 
IOP Publishing Ltd is not responsible for any errors or omissions 
in this version of the manuscript or any version derived from it. 
The Version of Record is available online at \url{https://doi.org/10.1088/1748-0221/15/03/C03055}.

\title{\boldmath Status of the PANDA Barrel DIRC}


\author[a,1]{C. Schwarz,\footnote{Corresponding author.}}
\author[a,b]{A.~Ali,}
\author[a]{A.~Belias,}
\author[a]{R.~Dzhygadlo,}
\author[a]{A.~Gerhardt,}
\author[a,b]{M.~Krebs,}
\author[a]{D.~Lehmann,}
\author[a,b]{K.~Peters,}
\author[a]{G.~Schepers,}
\author[a]{J.~Schwiening,}
\author[a]{M.~Traxler,}
\author[c]{L.~Schmitt,}
\author[d]{M.~B\"{o}hm,}
\author[d]{A.~Lehmann,}
\author[d]{M.~Pfaffinger,}
\author[d]{S.~Stelter,}
\author[d]{F.~Uhlig,}
\author[e]{M.~D\"{u}ren,}
\author[e]{E.~Etzelm\"{u}ller,}
\author[e]{K.~F\"{o}hl,}
\author[e]{A.~Hayrapetyan,}
\author[e]{I.~K\"{o}seoglu,}
\author[e]{K.~Kreutzfeld,}
\author[e]{J.~Rieke,}
\author[e]{M.~Schmidt,}
\author[e]{T.~Wasem,}
\author[f]{and C.~Sfienti}
\affil[a]{GSI Helmholtzzentrum f\"ur Schwerionenforschung GmbH, Darmstadt, Germany}
\affil[b]{Goethe University, Frankfurt a.M., Germany}
\affil[c]{FAIR, Facility for Antiproton and Ion Research in Europe, Darmstadt, Germany}
\affil[d]{Friedrich Alexander-University of Erlangen-Nuremberg, Erlangen, Germany}
\affil[e]{II. Physikalisches Institut, Justus Liebig-University of Giessen, Giessen, Germany}
\affil[f]{Institut f\"{u}r Kernphysik, Johannes Gutenberg-University of Mainz, Mainz, Germany}

\maketitle
\flushbottom

\abstract{The PANDA experiment will use cooled antiproton beams with high intensity stored
in the High Energy Storage Ring at FAIR. Reactions on a fixed target producing charmed hadrons 
will shed light on
the strong QCD.  
Three ring imaging Cherenkov counters are used for charged particle identification. 
The status of the Barrel DIRC (Detection of Internally Reflected Cherenkov light) is described.
Its design is robust and its performance validated in experiments with test beams. 
The PANDA Barrel DIRC has entered the construction phase and will be installed in 2023/2024.
}
\vspace{10mm}



\section{The PANDA Experiment at FAIR}
\label{sec:intro}

The new international FAIR accelerator complex  near GSI in Darmstadt, Germany, is currently under construction. 
The heart of the facility is an underground ring accelerator 
with a circumference of 1100 meters, which is being excavated. 
The existing GSI accelerator complex will serve as an injector for FAIR. The experimental 
pillars are NUSTAR, CBM, APPA, and PANDA (antiProton Annihilation at DArmstadt). 
The latter is using antiproton beams with unprecedented intensity and quality, stored and 
accelerated in the High Energy Storage Ring (HESR). The antiproton beam with the momentum range from 1.5 to 15 GeV/c
annihilates with a fixed target. The experiments will address questions of QCD 
in an energy region where perturbation theory is still valid, but the influence of strong QCD cannot
be neglected. 
The luminosity of up to $2 \cdot 10^{32} cm^{-2}s^{-1}$, and the momentum resolution of
the antiproton beam down to \mbox{$\Delta$p/p = 4$\cdot10^{-5}$} enable high precision 
spectroscopy. The energy range allows accessing resonances above the
threshold for open charm mesons. Thus, the detection of kaons plays an important role for 
the identification of the reaction channel. The PANDA detector consists of two parts, 
a hermetic target spectrometer and a forward spectrometer for polar angles up to  $5^\circ$ and 
$10^\circ$ in vertical and horizontal directions, respectively. The target spectrometer includes,
beside a tracking system, an electromagnetic lead tungstate calorimeter, a muon range system, and 
ring imaging Cherenkov counters for charged particle identification (PID).
The PID inside the calorimeter has to fulfill the detector requirements 
of limited space and limited mass.
Therefore, two Cherenkov counters using the DIRC principle were chosen. 
In a  DIRC detector the radiator acts also as a lightguide. 
The lightguide preserves the 
information of the Cherenkov angle of the charged particle over many internal photon reflections.
The Barrel DIRC
covers polar angles between $22^\circ$ and $140^\circ$ and will achieve a pion-kaon 
separation of 3 standard deviations (s.d.) up to 3.5 GeV/$c$.
At smaller polar angles an Endcap Disc DIRC \cite{Etzelmueller19} and  
a focusing aerogel RICH \cite{Kononov19} provide charged PID.
The focus of the following sections will be on the 
Barrel DIRC, its design, the results of the test beams, and selected prototype tests.

\section{The Design of the Barrel DIRC}
\label{sec:design}

\begin{figure}[tbh]
\captionsetup{width=0.8\textwidth}
\centering 
\includegraphics[width=.8\textwidth,trim=0 0 0 0,clip]{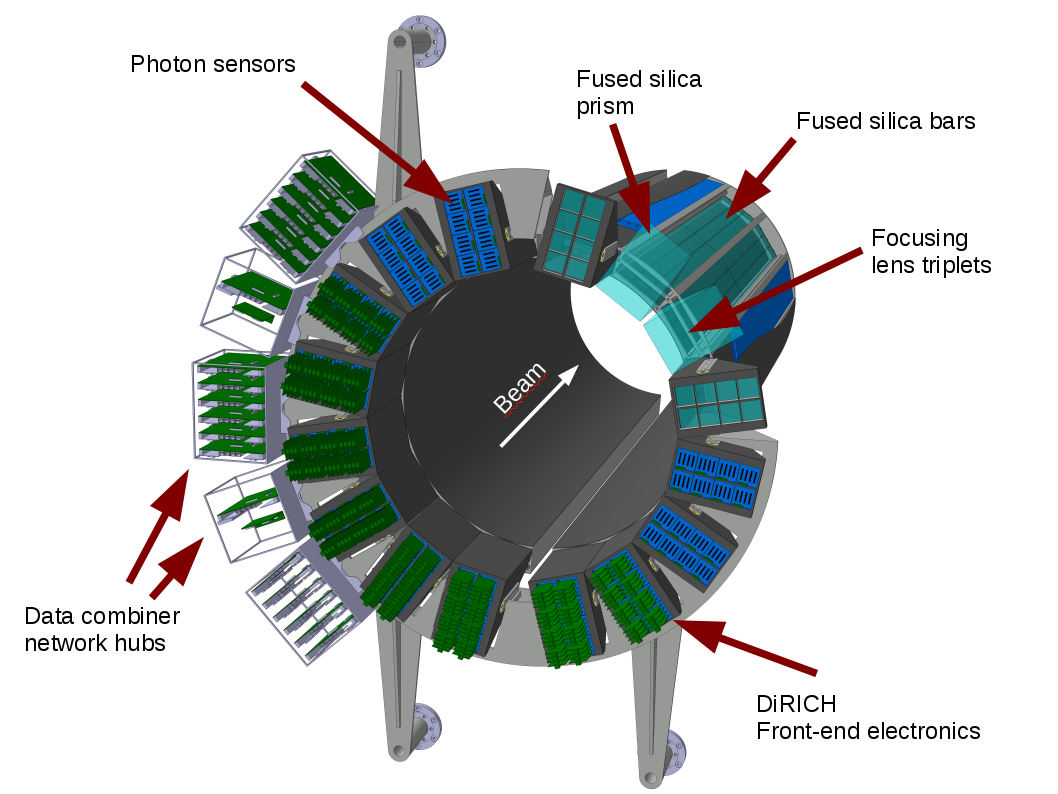}
\caption{\label{fig:1} Schematic of the PANDA Barrel DIRC.}
\end{figure}
The design of the PANDA Barrel DIRC (Figure \ref{fig:1}) is inspired by the successful BaBar DIRC and 
the R\&D for the SuperB FDIRC with innovative improvements.
The readout volume is a prism. It is smaller and made from synthetic fused silica
to minimize the influence of background radiation from the accelerator.
The use of pixelated Microchannel Plate Photomultiplier 
Tubes (MCP-PMTs) allows the operation of the readout within a magnetic field of 1 Tesla inside the
solenoid magnet of the target spectrometer. The fast detector response and the frontend electronics 
aim for a timing precision of
the readout chain of about 100 ps \cite{TDR,Schwiening18}. 

\begin{figure}[bth]
\captionsetup{width=0.8\textwidth}
\centering 
\includegraphics[width=.65\textwidth,trim=0 0 0 0,clip]{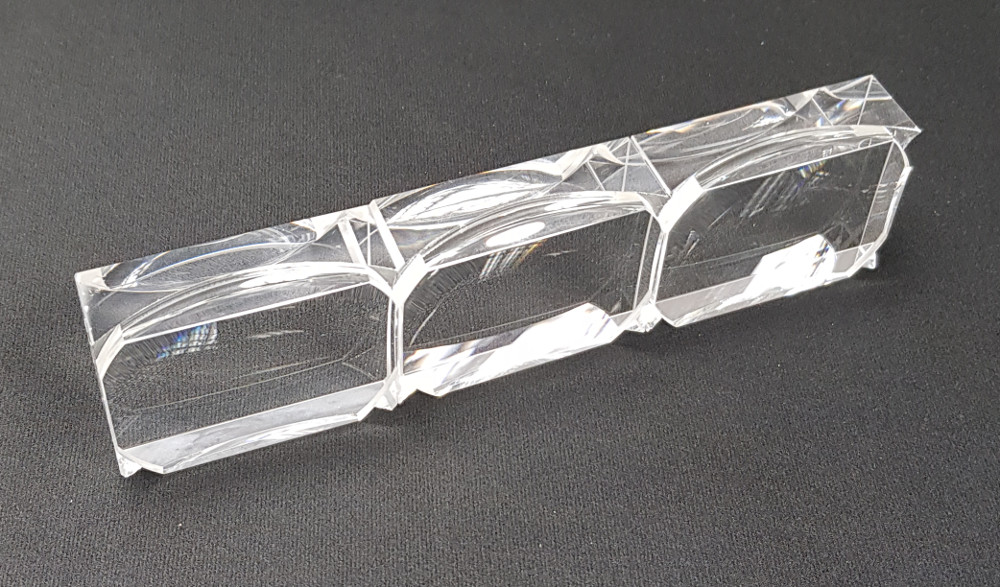}
\caption{\label{fig:3} Three focusing lens triplets, each a lanthanum crown glass lens between 
two fused silica pieces with
flat outer surfaces, focus the light from the radiator bar on the back side of the prism.}
\end{figure}

Focusing optics are needed due to the compact readout volume. 
The positioning of the photodetectors in the magnetic field and space 
limitations favored the use of a lens system. A picture of three lens triplets 
built by Befort Wetzlar Optics \cite{Befort} 
with the width of 53 mm are shown in Figure \ref{fig:3}.
The radiator barrel consists of 16 bar boxes containing three radiator bars each. The radiator bars 
are made from synthetic fused silica,  17 mm thick, 53 mm wide, and 2400 mm long. Two 1200 mm long 
pieces are 
glued end-to-end to form one radiator bar. 
With wider bars the needed number of entities in the barrel becomes smaller.  
The chosen width is the result of a cost optimization without deteriorating the required performance. 
At each downstream end a mirror is attached, reflecting the photons back towards the readout volume. 
A lens is glued to the upstream end of the bar and a RTV-silicone cookie couples the lens system
to a 300 mm-long
prism with an opening angle of $33^\circ$. 
An array of $2\times4$ MCP-PMTs, each with an $8\times8$ anode grid with a pixel pitch of $6.5$~mm, 
is coupled to the prism using a RTV-silicone cookie.
\section{Experiments with Test Beams}

The Barrel DIRC prototypes were tested in several experiments with  particle beams from 2008 to 2018. 
The measured photon number and single photon angular 
resolution of the setups were compared to predictions from GEANT4 \cite{geant4} simulations. 
The baseline design was validated in the campaign at CERN-PS in 2015. The following 
campaigns studied the possible reduction of the number of MCP-PMTs for cost optimization,
the performance of a wide radiator plate, 
different optical couplings, and housing and cable routing of the frontend electronics.
The prototypes consisted of a
single radiator bar or plate coupled to a prism with or without a focusing lens system and were read out by
up to 15 MCP-PMTs. 

In the beam tests MCP-PMTs like the XP85012/A1-Q from PHOTONIS \cite{photonis} were used.
The latest tubes are expected to survive 10 years of the operation of the PANDA experiment \cite{lehmann:mcp}.

The electronics for the readout of the MCP-PMTs in the prototype was based on the 
Trigger Readout Board Version 3 (TRB3) of the
HADES collaboration, which uses fast TDC channels implemented in FPGAs \cite{trb3-jinst}. 
The boards measure the time of
arrival and the time over threshold of logical signals coming from PADIWA discriminator boards
\cite{cardinali:padiwa} plugged onto the MCP-PMTs.

Between the radiator bar and the prism is the focusing lens system. Initial versions with a
focusing plano-convex lens on the radiator and an air gap between the prism and the lens allowed an easy 
separation of the expansion volume from the radiator for maintenance. 
However, many photons were reflected back by internal reflection from the lens surface.
In addition, the single refracting surface caused a parabolic-shaped focal plane. Therefore, the
air gap was filled with synthetic fused silica and the lens changed to a material
with a high refraction index and
sufficient transmission for UV-photons, lanthanum-crown glass LaK33.
The lens consists of a defocusing and a focusing surfaces to yield a sufficiently flat focal plane.
The transmission loss
of LaK33 at a wavelength of 420 nm is 1.3~\% for 1 Gy X-rays \cite{eRD14_18} 
and the anticipated radiation for the lens is 4 Gy within 10 years of operation \cite{TDR}. 

\begin{figure}[tbh]
\captionsetup{width=0.8\textwidth}
\centering 
\includegraphics[width=.7\textwidth,trim=0 0 0 0,clip]{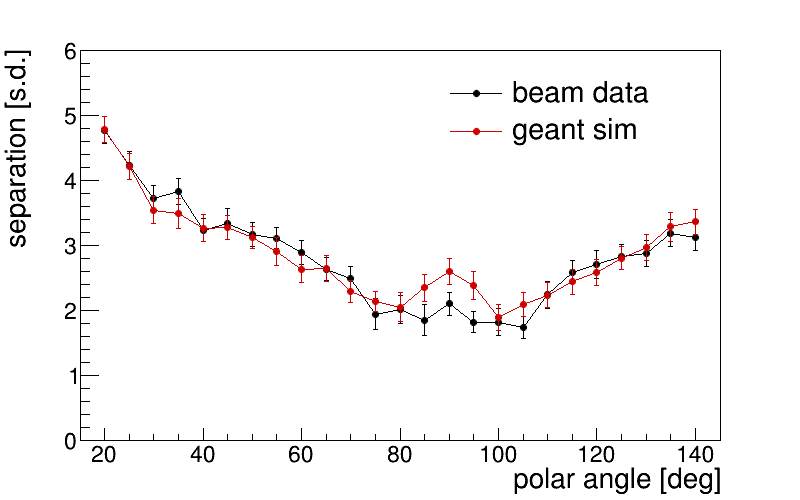}
\caption{\label{fig:2} Separation power for pion/proton at particle momenta of 7 GeV/c 
coming from the timing reconstruction method. This corresponds to
the separation power for pion/kaon at momenta of 3.5 GeV/c.}
\end{figure}
The geometry of the radiator, narrow bars or a single wide plate, requires different
reconstruction methods to identify the particle type \cite{Dzhygadlo19}.
The result of a maximum likelihood test using the time of arrival of photons for 
different particle types is
shown in Figure \ref{fig:2} for the test beam campaign in 2018 where pions and protons were 
cleanly tagged by a time-of-flight system. 

The separation between pions and protons with the momentum of $7$~GeV/c is equivalent to the 
separation between pions and kaons at $3.5$~GeV/c,
which is the designed upper momentum limit of the Barrel DIRC in PANDA.

\section{Finalizing the R\&D and Construction}
\begin{figure}[b]
\captionsetup{width=0.8\textwidth}
\centering 
\includegraphics[width=.615\textwidth,trim=0 0 0 0,clip]{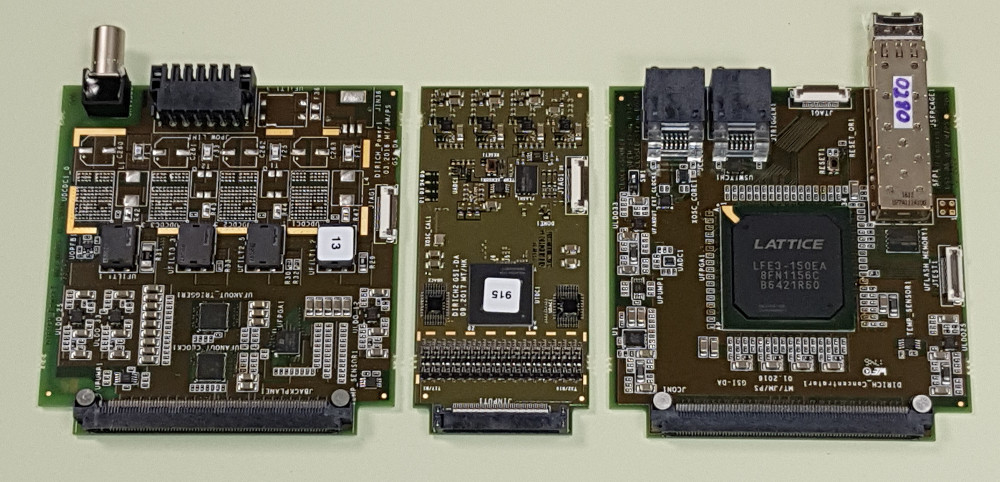}
\includegraphics[width=.35\textwidth,trim=0 0 0 0,clip]{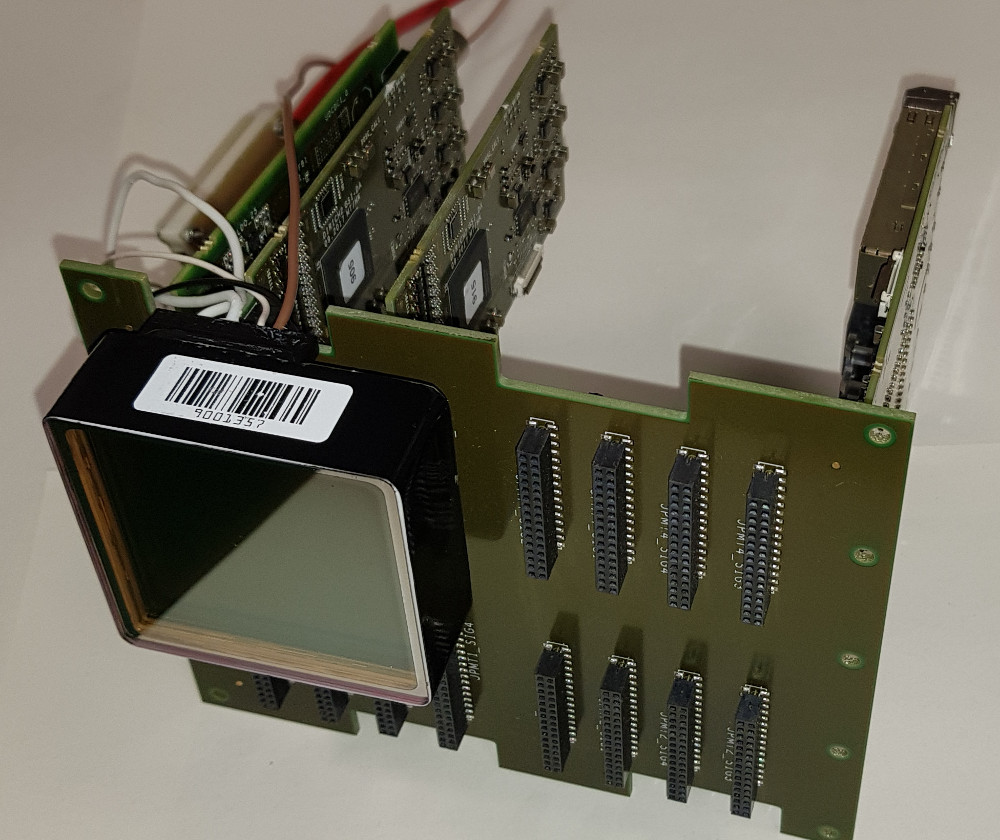}
\caption{\label{fig:x} The DiRICH system: a power card, a DiRICH card, and a data concentrator 
(left side) are plugged in a backplane together with the MCP-PMT (right side).}
\end{figure}
Following the successful performance validation with particle beams and the completion of 
the technical design report \cite{TDR}, the PANDA Barrel DIRC project has entered the construction phase.
The Barrel DIRC components with the longest production time, 
the radiators, have been ordered from Nikon \cite{nikon} and the tendering of the MCP-PMTs 
is at an advanced stage.

Several important R\&D topics are still being investigated.
The latest generation of the readout electronics, the DiRICH system \cite{traxler:dirich} comprises a backplane, 
a PC board with plugs for  MCP-PMTs on one side and three types of electronic cards on the other side (Figure \ref{fig:x}).
The electronic cards are 
a power supply, a data concentrator, and a DiRICH card. The latter contains 32 FPGA-based
discriminators and TDC channels. The discriminator input stage, originally designed for multianode PMTs, 
still needs to be adapted to the fast signals of the
MCP-PMTs. 

The coupling between the bar boxes and prisms and between the MCP-PMTs and the prism 
will be done with RTV-silicone cookies and is subject of
ongoing tests. Cookies with layers of different elasticity are produced to 
achieve the optimum contact between the optical surfaces. A firmer outer layer with a slightly convex
shape helps to prevent the appearance of air bubbles connecting the parts.

\begin{figure}[bh]
\captionsetup{width=0.8\textwidth}
\centering 
\includegraphics[width=.63\textwidth,trim=0 0 0 0,clip]{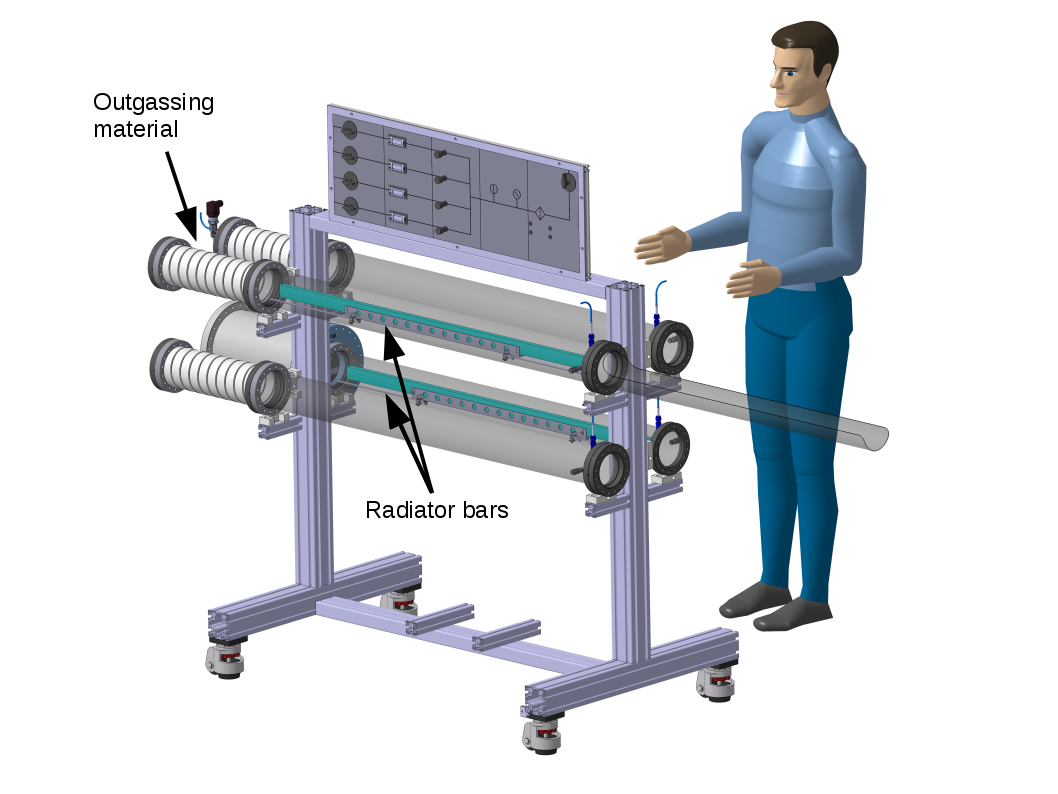}
\includegraphics[width=.34\textwidth,trim=0 0 0 0,clip]{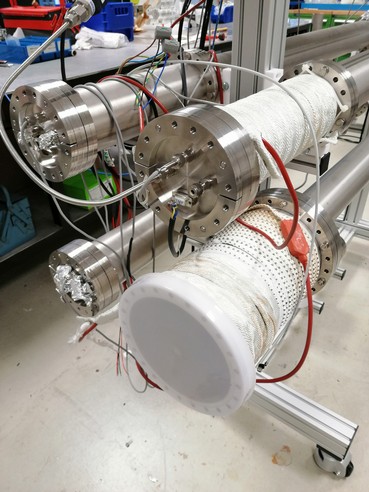}
\caption{\label{fig:4} Schematic of the setup for the measurement of the impact of outgassing 
on the bar surfaces (left side). 
The heatable volumes are connected to the tubes with the the radiator bars. 
The photograph shows the volumes for the outgassing materials (right side).  
}
\end{figure}

The material of the bar boxes can be aluminum or  carbon fiber reinforced polymer (CFRP). While the 
CFRP minimizes the material budget, the long-term outgassing of the material and its effect 
on the surfaces of the radiator are not well known.
Therefore, a setup was built to test the long-term outgassing behaviour of materials
which may be used in the construction of the bar boxes.
The radiator is sitting in a chemically cleaned stainless-steel tube with an attached steel 
container which contains the possibly pollutant material. 
A flow of nitrogen gas transports any possible outgassing pollutant across the surfaces of the radiator. 
The material to test
can be heated to accelerate the outgassing. 
The internal reflection coefficient of the possible polluted bar surface is 
measured at regular intervals, ranging from a few days to several weeks, 
with a laser scanning setup and compared to the reference bar.
The setup with four testing tubes is shown in Figure \ref{fig:4}.

With the arrival of the first radiator bars in 2020 the Barrel DIRC has entered the construction phase.
The experimental hall for the PANDA detector will be ready to move in and to install first basic elements,
like the solenoid, in 2022. The Barrel DIRC will then be installed in 2023/2024. 



\section*{Acknowledgments}

This work was supported by 
HGS-HIRe, 
HIC 
for FAIR, 
BNL, and eRD14.
We thank the GSI and CERN staff for the opportunity to use 
the beam facilities and for their on-site support.


\end{document}